\documentclass[prl,preprint,superscriptaddress]{revtex4}

\usepackage[dvipdfmx]{graphicx}
\usepackage{dcolumn}
\usepackage{bm}

\usepackage{color}

\begin{document}


\title{Non-trivial surface states of samarium hexaboride at the (111) surface}

\author{Yoshiyuki Ohtsubo}
\email{y_oh@fbs.osaka-u.ac.jp}
\affiliation{Graduate School of Frontier Biosciences, Osaka University, Suita 565-0871, Japan}
\affiliation{Department of Physics, Graduate School of Science, Osaka University, Toyonaka 560-0043, Japan}
\author{Yuki Yamashita}
\author{Kenta Hagiwara}
\affiliation{Department of Physics, Graduate School of Science, Osaka University, Toyonaka 560-0043, Japan}
\author{Shin-ichiro Ideta}
\author{Kiyohisa Tanaka}
\affiliation{Institute for Molecular Science, Okazaki 444-8585, Japan}
\author{Ryu Yukawa}
\author{Koji Horiba}
\author{Hiroshi Kumigashira}
\affiliation{Photon Factory, Institute of Materials Structure Science, High Energy Accelerator Research Organization (KEK), 1-1 Oho, Tsukuba 305-0801, Japan}
\author{Koji Miyamoto}
\author{Taichi Okuda}
\affiliation{HiSOR, Hiroshima University, Higashi-Hiroshima 739-0046, Japan}
\author{Wataru Hirano}
\author{Fumitoshi Iga}
\affiliation{College of Science, Ibaraki University, Mito 310-8512, Japan}
\author{Shin-ichi Kimura}
\email{kimura@fbs.osaka-u.ac.jp}
\affiliation{Graduate School of Frontier Biosciences, Osaka University, Suita 565-0871, Japan}
\affiliation{Department of Physics, Graduate School of Science, Osaka University, Toyonaka 560-0043, Japan}

\date{\today}

\begin{abstract}
The peculiar metallic electronic states observed in the Kondo insulator, samarium hexaboride (SmB$_6$), has stimulated considerable attention among those studying non-trivial electronic phenomena. However, experimental studies of these states have led to controversial conclusions mainly to the difficulty and inhomogeneity of the SmB$_6$ crystal surface. Here, we show the detailed electronic structure of SmB$_6$ with angle-resolved photoelectron spectroscopy measurements of the three-fold (111) surface where only two inequivalent time-reversal-invariant momenta (TRIM) exist. We observe the metallic two-dimensional state was dispersed across the bulk Kondo gap. Its helical in-plane spin polarisation around the surface TRIM suggests that SmB$_6$ is topologically non-trivial, according to the topological classification theory for weakly correlated systems. Based on these results, we propose a simple picture of the controversial topological classification of SmB$_6$.
\end{abstract}

\maketitle

\section*{Introduction}
The coexistence of strong electron correlation and topological order is garnering much attention nowadays because of various peculiar electronic phenomena that are driven by their synergetic effect \cite{Dzero10, Pesin10, Allen16}.
The strong topological insulator realised in the bulk (3D) Kondo insulator, namely the topological Kondo insulator (TKI) \cite{Dzero10}, is being extensively considered as a suitable field to study these effects such as non-trivial reconstruction of the topological surface states (TSS) due to electron correlation \cite{Alexandrov15, Peters16} and spin collective excitation, which can break the TSS without time-reversal symmetry breakdown \cite{Kapil15}.

Samarium hexaboride (SmB$_6$) is a long-known Kondo insulator, which opens the bulk bandgap at low temperature because of the Kondo effect \cite{Cohen70}. It is the first material proposed as a candidate for TKI, which hosts metallic TSS coexisting with strong electron correlation \cite{Dzero10, Takimoto11}.
To investigate this unconventional TSS, extensive studies that focused on its surface electronic structure were performed \cite{Miyazaki12, Xu13, Neupane13, Jiang13, Xu14, Hlawenka15} mainly by using angle-resolved photoelectron spectroscopy (ARPES) and spin-resolved ARPES (SARPES) on the cleaved (001) surface of SmB$_6$.
Although the metallic surface states dispersed across the bulk Kondo gap were discovered in TKI, as predicted \cite{Miyazaki12, Xu13, Neupane13, Jiang13, Xu14}, a subsequent high-resolution ARPES study made a counter-claim regarding such TKI assignment by stating that some of the metallic surface states do not disperse continuously across the bulk Kondo gap but accidentally lie at the Fermi level ($E_{\rm F}$) \cite{Hlawenka15}.
Although numerous other studies such as surface-transport \cite{Phelan14} and scanning tunnelling microscopy \cite{Rosler14, Miyamachi17} strongly suggest the topologically non-trivial nature of SmB$_6$, the detailed surface electronic and spin texture of SmB$_6$ have remained unclear because of this disagreement.
Moreover, a peculiar Fermi surface behaviour of SmB$_6$ has been reported recently through the de Haas--van Alphen (dHvA) measurements \cite{Li14, Tan15, Hartstein17, Xiang17}.
All groups reported carriers lying at $E_{\rm F}$ without electrical conduction, but its interpretation, 2D \cite{Li14, Xiang17} or 3D \cite{Tan15, Hartstein17}, is still under debate.
Because of these background, it is desirable to elucidate the surface electronic structure of SmB$_6$ and its topological order.

In this work, we report the topological surface state of a typical candidate for TKI, SmB$_6$, which is observed on the three-fold (111) surface by ARPES.
We can determine the topological order on the SmB$_6$(111) surface from the surface Fermi contours because of the smaller number of inequivalent surface time-reversal invariant momenta (TRIM) and the absence of commensurate and long-range surface reconstructions, as reported for the (001) surfaces \cite{Miyazaki12, Hlawenka15, Rosler14}.
The metallic two-dimensional state is clearly observed as dispersed across the bulk Kondo gap opening around the Fermi level at low temperature.
Its helical in-plane spin polarisation around the $\bar{\rm M}$ point of the surface Brillouin zone, which is one of the surface TRIMs, indicates a non-trivial topological order of SmB$_6$.
Based on these results, we propose a simple picture of topological-insulating SmB$_6$.

\section*{Results}
\subsection*{A (111) surface of SmB$_6$}
One of the difficulties in determining the detailed surface electronic structure of SmB$_6$ from the ARPES results is its rather complex surface TRIM conformation on the (001) surface.
As shown in Fig. 1 (a), there are three inequivalent surface TRIMs on SmB$_6$(001). 
While the TSS should appear as an odd number of closed Fermi contours (FC) enclosing the TRIMs an odd number of times, three such inequivalent TRIMs allow various possibilities regarding the appearance of the TSS \cite{Teo08}.
Considering the multiple surface terminations on the cleaved (001) surface \cite{Rosler14}, it is quite a difficult problem to determine the topological order of SmB$_6$ solely from the electronic structure of the (001) surface.
To overcome this problem, the surface electronic structure with a different surface orientation is desired.
However, the SmB$_6$ single crystal can be cleaved only along (001). Hence, almost no studies have been performed so far on the surface electronic structures with different orientations.
Only one set of ARPES data taken from the (110) surface prepared by a similar method to ours has been provided as a preprint \cite{Denlinger16}, but the (110) plane has the same problem as (001); it also contains multiple inequivalent surface TRIMs.
The (111) surface of SmB$_6$ is a promising orientation for determining its topological order because there are only two inequivalent TRIMs (right panel of Fig. 1 (a)): one $\bar{\Gamma}$ and three equivalent $\bar{\rm M}$.
Note that the other high-symmetry point $\bar{\rm K}$ is not a TRIM.
With this simple surface-TRIM conformation, the TSS must appear around one surface TRIM, and thus the determination of the topological order becomes very easy when compared with the previous case.
However, no work on the surface electronic structure of SmB$_6$(111) has been reported so far.

In order to obtain the SmB$_6$(111) clean surface, we heated the single crystal up to 1700$\pm$30 K for 15 min in an ultra-high vacuum chambers by using the same method as applied for YbB$_{12}$(001) \cite{Hagiwara16, Hagiwara17}.
After heating the sample, one can see sharp and low-background low-energy electron diffraction (LEED) pattern, as shown in Fig. 1 (b). The three-fold triangular lattice shown by the LEED pattern is consistent with the (111) surface truncated from the simple-cubic lattice (see Fig. 1 (c)).
 Faint streaks between the integer-order diffraction spots are also seen in the LEED pattern. They would be due to the small area of the facets or long-range surface superstructures without wide commensurate surface areas.
It should be noted that the topological order of the material is not influenced by such disordered surface structures and we observed no electronic states related to such surface superstructures in the 1st surface Brillouin zone (SBZ), as discussed in the following sections.
The (111) surface obtained by this method would be terminated by the Boron clusters, according to the angle-integrated photoelectron spectroscopy \cite{Ohtsubo18}. 

\subsection*{Surface electronic structure of SmB$_6$(111)}
Figure 2 (a) shows the FC around $E_{\rm F}$ measured with circularly polarised photons at 35 eV.
The spectra obtained by using both right- and left-handed polarisations are summed up to avoid circular dichroism.
It clearly shows the deformed hexagonal FC enclosing the centre of the SBZ, which is the $\bar{\Gamma}$ point ($k_{\rm x}$ = $k_{\rm y}$ = 0 \AA$^{-1}$).
From the symmetrised wide-range overview shown in Fig. 2 (b), one can find that the deformed hexagon is a part of the ovals enclosing the $\bar{\rm M}$ points, as indicated by the dashed guide.
Around $E_{\rm F}$, no other states are observed by ARPES, indicating that the long-range surface structures observed as faint facets in the LEED pattern (Fig. 1 (b)) play no major role for the surface states around $E_{\rm F}$.
The size of the FCs observed here might be related to the peculiar Fermi surfaces obtained by dHvA measurements \cite{Li14, Tan15, Hartstein17, Xiang17}.
For the sake of comparison, we evaluated the sizes of the FCs in the supplementary note 5.

Figure 3 (a) shows the band dispersions along $\bar{\Gamma}$--$\bar{\rm M}$ ([11$\bar{2}$]).
In order to trace the band dispersion, we took the momentum and energy distribution curves as shown in Figs. 3 (b) and 3 (c), respectively.
The peak positions are plotted with the guides in Fig. 3 (d) so that they could be compared with the 2D data in Fig. 3 (a).
From the MDCs, the highly dispersive bands, $S1$ and $S2$ in Fig. 3 (d), are clearly observed as the peaks.
From the EDCs, less dispersive bands, $F$ at $\sim$0.03 eV and the other, underlying band at $\sim$ 0.17 eV are observed as the peaks.
In addition, the highly dispersive bands $S1$ and $S2$ appears in EDCs as broad humps, as indicated by the open triangles in Fig. 3 (c).
Although it is difficult to determine the strict peak positions in the EDCs, the energy region indicated by the bars in Figs. 3 (c) and 3 (d) have higher intensities than the other EDC spectra (the overlap of them are shown in supplementary note 4).

The band lying at the Fermi level, $S1$, is independent of the incident photon energy range of 15-39 eV, indicating the two-dimensionality from the surface origin.
On the left side of Fig. 3 (d), the projected bulk bands based on the theoretical calculation in ref. \cite{Baruselli16} is shown as the shaded area.
Comparing this with the ARPES data, $S1$ is out of the projected bulk bands and hence it should be the surface-state band.

For the other band, $S2$, which is mostly in the projected bulk bands, it is difficult to conclude whether it comes from surface or bulk in the photon-energy range, in which $S2$ is observable.
In this article, we don't conclude the origin of $S2$, from the surface resonance \cite{Appelbaum76} or bulk Sm-5d bands.
The detailed dataset and its analysis is shown in supplementary note 2.
From the EDCs, it is shown that the $F$ band appears separately from $S1$.
However, it is also difficult to conclude the origin of the $F$ band from the ARPES data.
While it appears irrespective to the incident photon energies, the bulk counterpart, the Sm-4$f$ band, is nearly localized and thus it should also show almost no dispersion along the surface normal.
The upper edge of the bulk projection in Fig. 3 (d) is slightly lower than $F$, but the exact values from theoretical calculations, such as the size of the bandgap and the position of the Fermi level, does not always agree with those from experiments.
Therefore, the origin of $F$ is not clear from the spin-integrated ARPES.
The same analysis was also performed along $\bar{\Gamma}-\bar{\rm K}$ and the similar states to $S1$, $S2$ and $F$ were found (see supplementary note 3).

The $S1$ and $S2$ change their slopes drastically around the crossing point ($\sim\pm$0.2 \AA$^{-1}$) with $F$.
These hybridisations are probably driven by the Kondo effect between localised Sm-4$f$ and itinerant Sm-5$d$ states.
The upper band $S1$ clearly disperses across $E_{\rm F}$ and this band forms the oval FC observed in Fig. 2.
The dispersions of the surface state observed here agree well with the expected behaviour of topological surface states, namely, continuous dispersion across the bulk bandgap and closed FCs around the surface TRIMs.
Then, we performed SARPES measurements to examine the spin texture of the FCs, which is regarded as one of the clearest evidence of the topological order of the material.

\subsection*{Spin texture of the SmB$_6$(111) surface states}
Figure 4 (a) shows the spin-resolved energy distribution curves (EDCs) around the Fermi level measured along $\bar{\Gamma}$--$\bar{\rm M}$ at 20 K.
The spin polarisation along [$\bar{1}$10] and the in-plane orientation perpendicular to $k_{{\rm y}//[11\bar{2}]}$ were resolved.
From the EDC Cuts 1 to 4, one can easily find that the spin-polarised feature towards [1$\bar{1}$0], indicated by the negative spin polarisation values, disperses from +0.03 to -0.03 eV across $E_{\rm F}$, which is consistent with the metallic dispersion of the surface band $S1$.
At the opposite side of the SBZ (Cut 5 in Fig. 4 (a)), the opposite spin polarisation towards [$\bar{1}$10] (positive spin polarisation) is also observed.
Such spin inversion according to the sign inversion of $k_{\rm y}$ indicates that these spin polarisations conserve the time-reversal inversion symmetry.
The spin polarisation value at positive $k_{\rm y}$ is nearly twice that at negative $k_{\rm y}$. This would be due to the lack of mirror plane normal to $\bar{\Gamma}$--$\bar{\rm M}$. possibly because there is no mirror plane normal to $\bar{\Gamma}$--$\bar{\rm M}$, as shown in the atomic structure and the LEED pattern in Figs. 1 (c) and 1 (b), respectively.

One may doubt that if the energy resolution of the current SARPES setup, $\sim$30 meV, is enough to trace the spin polarisation of the surface states or not, since this resolution is close to the total size of the energy window where $S1$ is visible.
However, the SARPES data showed the clear spin polarisations well above the noise level evaluated by the standard statistical errors and evident spin-polarized peaks consistent with the dispersion of $S1$ obtained from the spin-integrated data (Figs. 2 and 3).
They proved that the current SARPES data is enough to trace the spin polarisation of the surface states,  without any ambiguity.

The spin-resolved EDCs indicate that the spin polarisations of the deeper features, from 0.15 eV at Cut 1 ($k_{\rm y}$ = +0.29 \AA$^{-1}$) to 0.04 eV at Cut 4 ($k_{\rm y}$ = +0.17 \AA$^{-1}$), are opposite to $S1$.
These deeper features correspond to $F$ and $S2$ observed by the spin-integrated ARPES.
The similar feature was also observed in the opposite side of the SBZ (Cut 5: $k_{\rm y}$ = -0.25 \AA$^{-1}$).
These polarisation values are also above the estimated errors as shown by the error bars in Fig. 4 (a).
If one assume the $S2$ and $F$ to be the surface bands, such spin polarizations could be understood as a result of space inversion asymmetry in the surface layers.
On the other hand, even if $S2$ and $F$ come from bulk bands, such spin polarization can appear because of the spin-dependent reflectivity of the Bloch waves at the surface \cite{Kimura10}.
Therefore, we can neither determine the origin of $S2$ and $F$, from surface or bulk, by the spin resolution.
From the dispersions of $S1$ and $F$, they apparently degenerate with each other close to $\bar{\rm M}$.
This behaviour might be the Kramers degeneracy between $S1$ and $F$ as the surface bands, or the surface band $S1$ merging into bulk bands $F$.
Anyway, to verify these assumptions,
dispersions of them in the vicinity of $\bar{\rm M}$ with spin resolution should be measured.
Such measurement was not possible in this work because of the limited energy resolution of SARPES; even far away from $\bar{\rm M}$ (around 0.4 \AA$^{-1}$), it is impossible to resolve $S1$ and $F$.
To examine this assumption, the higher energy resolution in SARPES is desirable.

The spin polarizations of the surface states along $\bar{\Gamma}$--$\bar{\rm K}$ were measured by the spin-resolved MDCs as shown in Fig. 4 (b).
The MDC peak heights along $\bar{\Gamma}$--$\bar{\rm K}$ are nearly equal to each other, reflecting the presence of the mirror plane normal to $\bar{\Gamma}$--$\bar{\rm K}$.
Although the MDC peak shape is not symmetric along $\bar{\Gamma}$--$\bar{\rm K}$, one can find its peaks at $k_{\rm x} \sim \pm 0.3$\AA$^{-1}$, which is consistent with those of metallic oval FCs shown in Fig. 2.
The origin of asymmetric peak shape and its influence to spin polarization is shown in supplementary discussion 2. 
The in-plane spin polarisations along $\bar{\Gamma}$--$\bar{\rm K}$ are consistent with those shown in Fig. 4 (a).
In addition, clear out-of-plane spin polarisations are also observed along $\bar{\Gamma}$--$\bar{\rm K}$, as shown in the lower part of Fig. 4 (b), whereas such spin polarisations are almost negligible along $\bar{\Gamma}$--$\bar{\rm M}$ (see supplementary note 6).
Along $\bar{\Gamma}$--$\bar{\rm K}$, the in-plane and out-of-plane spin polarisations of the photoelectrons are nearly equal to each other.
Such out-of-plane spin polarisations would be due to the coupling between spin-orbit interaction and the valley degree of freedom around $\bar{\rm K}$, where three-fold rotation symmetry appears without time-reversal symmetry, as observed in Tl/Si(111) \cite{Sakamoto09} and transition-metal dichalcogenides \cite{Suzuki14}.

Based on the SARPES spectra, we depicted a schematic drawing of the spin texture of the metallic surface states on SmB$_6$(111) in Fig. 4 (d).
As shown in the figure, the oval FC enclosing $\bar{\rm M}$ has clockwise spin polarisations along the in-plane orientations and finite out-of-plane ones away from the surface mirror plane parallel to $\bar{\Gamma}-\bar{\rm M}$.
Such non-zero component along the out-of-plane orientations is natural for topological states on the surfaces with three-fold rotational symmetry, $e.g.$ those on Bi$_2$Te$_3$ \cite{Souma11}.
The detailed discussion about the role of surface symmetry operations to the spin polarisations is shown in supplementary discussion 1.

The whole spin texture, both in-plane and out-of-plane ones, qualitatively agrees well with a recent theoretical calculation \cite{Baruselli16}, supposing the negative winding number $w$ = -1.
One should be careful for such comparison between theory and SARPES data, because sometimes spin polarisation of photoelectrons occurs artificially due to photoexcitation process and/or spin-orbital entanglement \cite{Henk95, Yaji17}.
However, it should be noted that the spin polarisation whose sign inverts with respect to time inversion, as shown in Fig. 4, cannot appear from spin-degenerate states even if such artificial spin polarisations occurred.
Although it is sometimes shortcoming to connect the observed spin polarisation of the photoelectrons to those of the initial states directly, it is evident that the initial states, $S1$, $S2$ and $F$, are somehow spin-polarized and its sign inverts according to the surface symmetry. This information is enough to discuss the topological order of the sample from its surface states, as shown in the following.

\section*{Discussion}
Based on the spin-polarisation and shape of the FCs, the topological order of SmB$_6$ is calculated.
In order to obtain the topological order of a material from its surface states, one has to obey the following procedure \cite{Teo08}: 

\noindent (i) Count up the FCs enclosing surface TRIMs.

\noindent (ii) Observe them by SARPES to check if they are spin polarized or not. The number counted in (i) is doubled for the spin degenerate states.

\noindent (iii) Examine the summed number. If it is odd, then the sample is a (strong) topological insulator. If even, the sample is normal, trivial insulator.

\noindent On SmB$_6$(111), the FCs enclosing $\bar{\rm M}$ appears three times (i); note that there are only three (not six) inequivalent $\bar{\rm M}$ points because of the translational symmetry by the surface reciprocal lattice vectors.
Since all the FCs here are spin polarized (ii), the total count in this case is three, the odd number.
According to this calculation, SmB$_6$ is determined to be a topological insulator, without any ambiguity.
Note that the same procedure is difficult to be applied to the (001) surface of SmB$_6$, since the number of FCs are still under discussion \cite{Xu13, Neupane13, Jiang13, Xu14, Hlawenka15}.
At first glance, this conclusion appears to conflict against a recent high-resolution ARPES data on (001) \cite{Hlawenka15}. However, they could be reconciled by an interpretation of the FCs observed on (001). Detailed discussion on this point is shown in supplementary discussion 3.
The other important point on the topological classification discussed above is that the detailed origin of the surface states. 
It is because the topological classification is merely from the number of the spin-polarized FCs \cite{Teo08, Kane10}.
In other words, if the odd-number of FCs were made by surface states derived from many-body resonance for example, the insulating substrate would be nothing but a topological insulator.
{
As a supplementary information, we discuss the possible origins of the surface states observed in this work (supplementary discussion 4).
}

The dispersion of TSS ($S1$) in this work does not show any Dirac point.
Instead, it shows quite a light velocity of $\sim$0.8 eV \AA (see supplementary note 4 for its estimation) only around the Fermi wavevector ($k_{\rm F}$).
Away from $k_{\rm F}$, $S1$ becomes quite heavy with almost no dispersion at the binding energy of $\sim$30 meV.
Such behaviour agrees well with the theoretically expected TSS dispersion modified by the Kondo breakdown \cite{Alexandrov15}.
The expected Fermi velocity in ref. \cite{Alexandrov15} is $\sim$0.3 eV \AA, showing an agreement of the order of magnitude with the experimental value above.
Further theoretical work taking the large size of FC overlapping with each other and/or out-of-plane spin components into account might be applicable to solve this factor 2-3 discrepancy.

{
At last, we'd like to state the limitation of the current work.
The topological classification procedure \cite{Teo08} assumes weakly correlated insulator.
Therefore, we cannot exclude a possible violation of such simple topological classification by strong electron correlation.
To the best of our knowledge, such work has never been published so far.
However, once such discovery has been achieved, the topological classification performed in this work should be revisited.
The other limitation is the bulk electronic structure of SmB$_6$ at 15-20 K, where we made the measurements.
It is commonly regarded that the rather wide activation gap ($\sim$20 meV) opens and the bulk electronic structure transforms to an insulator in this temperature range \cite{Flachbart01}.
However, a recent ARPES study \cite{Hlawenka15} has claimed that the bulk gap is still closed there, attributing the other small gap (3-5 meV, which opens below 10 K) is the real gap.
Although no experimental data supporting the claim above, the bulk band surviving at $E_{\rm F}$ around 20 K, has been reported so far, we have to admit that the topological classification in this work becomes nonsense in this temperature range, if this claim is correct.
Note that the claim in ref. \cite{Hlawenka15} is not the remnant bulk carriers thermally excited across the gap ($\sim$20 meV) but the firm band without any gap across the Fermi level.
It should also be noted that the peculiar Fermi surface observed by quantum oscillation experiments \cite{Li14, Tan15, Hartstein17, Xiang17} are not relevant to this possibility, because they were performed in much lower temperature range.

}

In summary, the topological surface state of a topological Kondo insulator candidate, SmB$_6$, was clarified with regard to the different surface orientation from the earlier works, the three-fold (111) surface, by angle-resolved photoelectron spectroscopy (ARPES), in this work.
The metallic two-dimensional state was clearly observed as dispersed across the bulk Kondo gap opening at the Fermi level at low temperature.
Its helical in-plane spin polarisation around the $\bar{\rm M}$ point of the surface Brillouin zone, which is one of the surface time-reversal invariant momenta, provided { the} evidence of the non-trivial topological order of SmB$_6$.
{ Based on these results, we propose a simple picture of topological-insulating SmB$_6$ to be} a fascinating groundwork to study peculiar electronic phenomena such as the synergetic effects { with} strong electron correlation.

\subsection*{Methods}
\subsection*{Sample preparation}
Single crystalline SmB$_6$ were grown by the floating-zone method by using an image furnace with four xenon lamps \cite{Iga98, Ellguth16}.
The sample cut along the (111) plane was mechanically polished in air until a mirror-like shiny surface was obtained with only a few scratches when observed under an optical microscope (multiple 10x magnification).

\subsection{ARPES and SARPES experimental setup}
The ARPES measurements were performed with synchrotron radiation at BL7U SAMRAI \cite{SAMRAI} of UVSOR-III and BL-2A MUSASHI of the Photon Factory. The photon energies used in these measurements ranged from 18 to 80 eV.
SARPES measurements were performed at HiSOR BL9B ESPRESSO \cite{Okuda11} with linearly polarised photons at 26 eV so that the photoelectron spin polarisation due to the circularly polarized photons are excluded \cite{Suga14}.
A pair of the very low energy electron diffraction (VLEED) detectors enable the three-dimensional detection of the spin polarizations \cite{Okuda15}.
The effective Sherman function of the spin detector was set to 0.3 and the acceptance angle for the detector was $\pm$1.5$^{\circ}$.
The energy resolutions of the spin-integrated and SARPES in this work were $\sim$15 and $\sim$30 meV, respectively.
The energy resolutions and photoelectron kinetic energies at the Fermi level $E_{\rm F}$ were calibrated using the Fermi edge of the photoelectron spectra from Ta foils attached to the samples.
The detailed experimental geometries are shown in supplementary note 1.

The SARPES spectra were measured four times as $I_p^1$, $I_n^1$, $I_n^2$, $I_p^2$ with this order, where $I_p^i$ and $I_n^i$ ($i$ = 1, 2) are the raw spectra obtained from the VLEED detector with positive and negative target magnetization, respectively.
Then, we got $I_p$ = $I_p^1$ + $I_p^2$ and $I_n$ = $I_n^1$ + $I_n^2$.
By this procedure, we compensate the time-dependent degradation of the surface states as well as the decay of incident photon flux (proportional to the beam current of the storage ring of HiSOR).
The spin polarisation of the SARPES spectra is calculated by $P$ = ($I_p$ - $I_n$)/$S_{eff}$ ($I_p$ + $I_n$), where $P$ is the spin polarisation shown in Fig. 4 and $S_{eff}$ the effective Sherman function. $S_{eff}$ is calibrated by the spin polarisation of the well-known surface state \cite{Okuda11}. The errors of $P$ is calculated as the standard statistical error. Then, the spin-resolved spectrum $I_{\pm}$, which are shown in Fig. 4, is calculated by $I_{\pm}$ = ($I_p$ + $I_n$) (1$\pm P$)/2.
For the SAPRES spectra in Fig. 4, no normalization nor smoothing procedures have been applied.

\subsection*{Data availability}
The datasets generated during and/or analysed during the current study are available from the corresponding author on reasonable request.

\subsection*{Ackowledgements}
We thank T. Nakamura, Y. Takeno and C. Wang for their support during general experiments.
We acknowledge S. Wu for his support during the experiments on BL9B at HiSOR.
Part of the ARPES measurements were performed under UVSOR proposals 29-553 and 30-577, Photon Factory proposals Nos. 2015G540 and 2017G537, and HiSOR proposal  No. 17AG017.
This work was also supported by the JSPS KAKENHI (Grants Nos. JP15H03676, JP17K18757, and JP23244066).

\subsection*{Author contributions}
Y.O., Y.Y. and K.H. conducted the ARPES experiments with assistance from S.-i.I., K.T., R.Y., K.H. and H.K. as well as the spin-resolved APRES assisted by K.M. and T.O..
W.H. and F.I. grew the single-crystal samples.
Y.O. and S.-i.K. wrote the text and were responsible for the overall direction of the research project.
All authors contributed to the scientific planning and discussions.

\subsection*{Competing Interests}
The authors declare no competing financial or non-financial interests.

\begin{figure}[p]
\includegraphics[width=80mm]{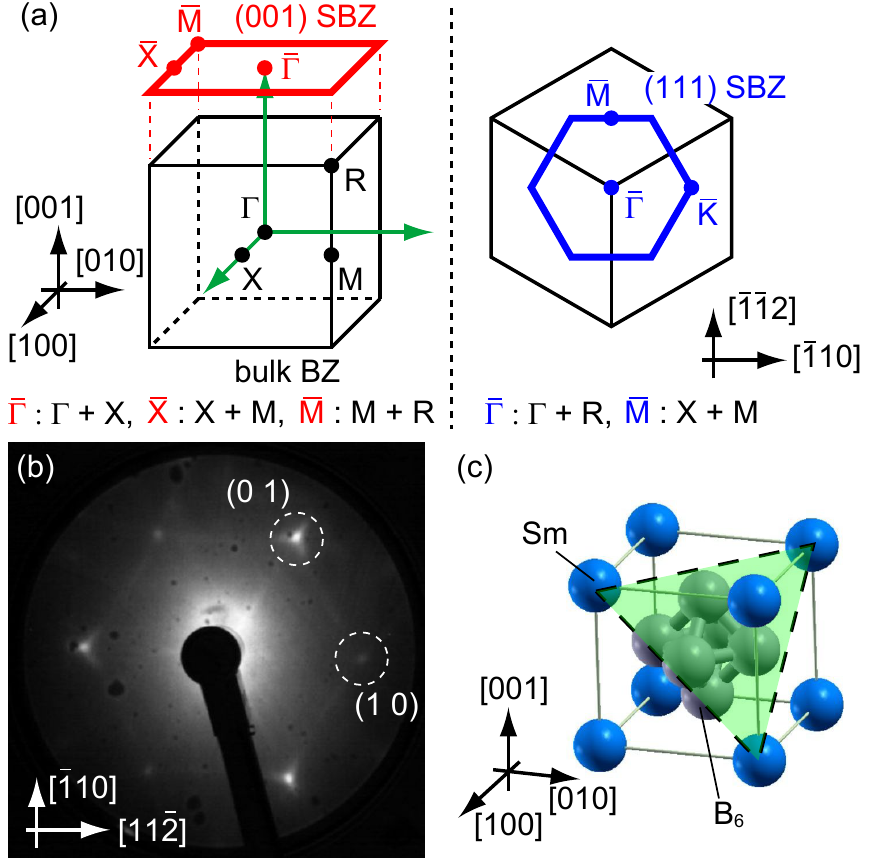}
\caption{\label{fig1} {\bf Atomic structure of SmB$_6$ and characteristics of the (111) surface} 
(a) Schematic drawings of the Brillouin zones (BZ). Thin (black) cubes are the 3D bulk BZ with time-reversal invariant momenta (TRIMs), and the thick (red and blue) lines are the first zone boundaries of the 2D surface Brillouin zones (SBZ) with the surface TRIMs.
(b) A low-energy electron diffraction (LEED) pattern of SmB$_6$(111) at room temperature. $E_{\rm P}$ = 22 eV.
(c) Crystal structure of the SmB$_6$. The dashed triangle indicates the (111) plane.
}
\end{figure}

\begin{figure}[p]
\includegraphics[width=80mm]{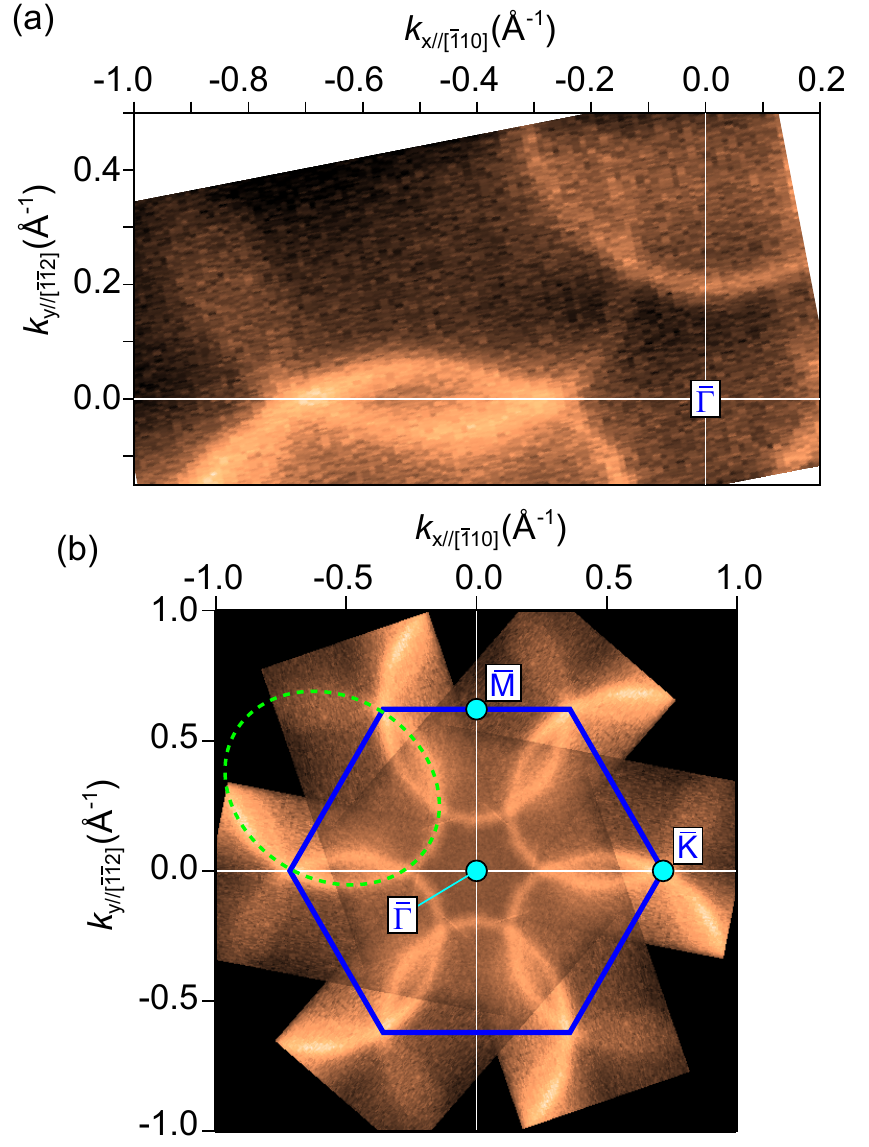}
\caption{\label{fig2} {\bf Fermi contour obtained by Angle-resolved photoelectron spectroscopy (ARPES)} 
The ARPES data were taken with circularly polarised photons ($h\nu$ = 35 eV) at 15 K. The ARPES intensities from left- and right-handed polarisations are summed up to show all the states without any influence of circular dichroism. The photon incident plane is slightly shifted from (11$\bar{2}$) because of small misalignment and angle sweep performed for the ARPES scan. This shift is smaller than 15$^{\circ}$.
(a) Fermi contour with an energy window of 10 meV.
(b) Symmetrised Fermi contour based on the three-fold rotation symmetry and time-reversal symmetry. A thick (blue) hexagon is the SBZ boundary of the (1$\times$1) surface unit cell.
}
\end{figure}

\begin{figure*}[p]
\includegraphics[width=150mm]{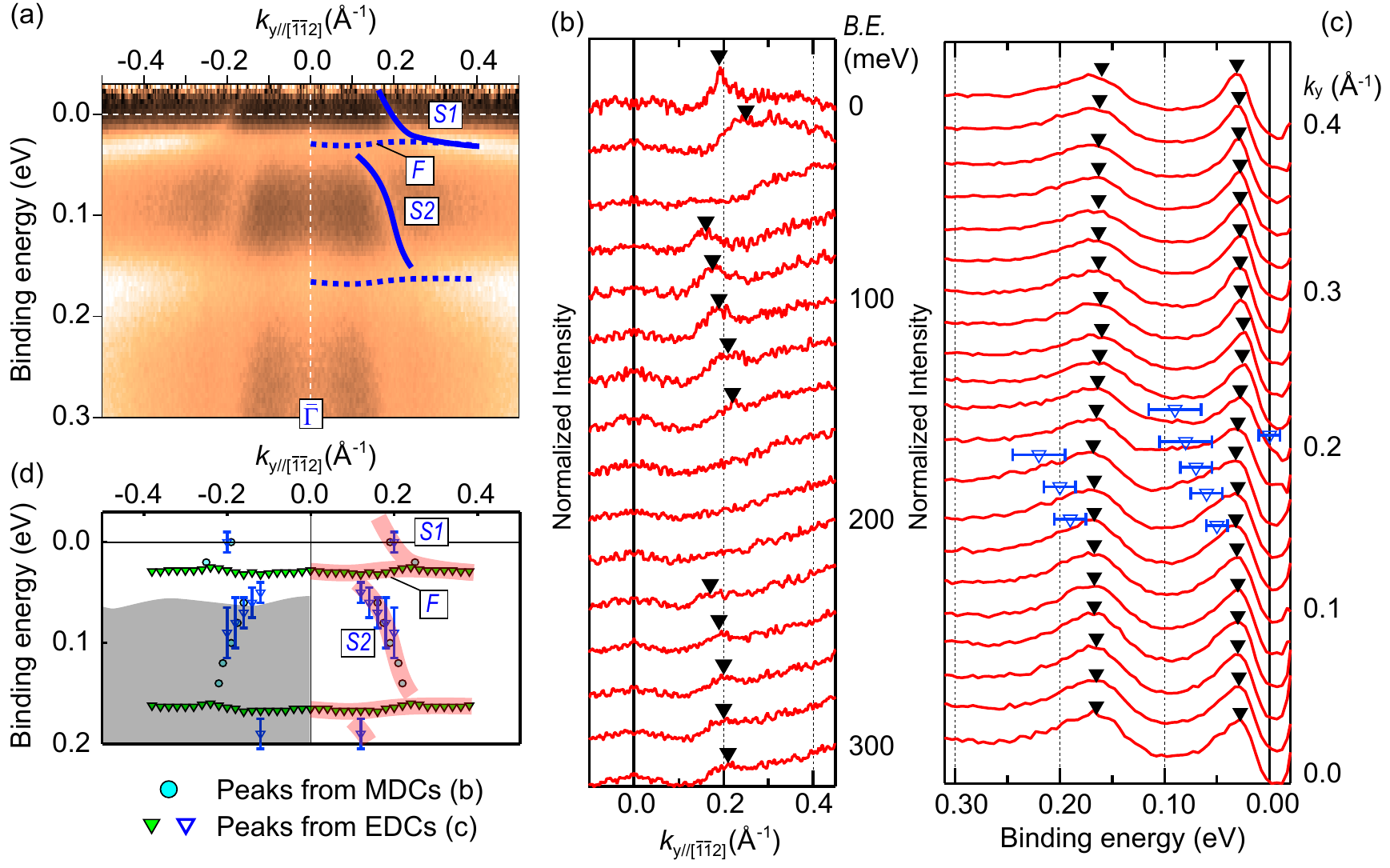}
\caption{\label{fig3} {\bf Band dispersions of SmB$_6$(111) around the Fermi level} 
ARPES data were taken with the same condition as Fig. 2.
(a) ARPES intensity plots along $\bar{\Gamma}$--$\bar{\rm M}$ symmetrised with respect to $\bar{\Gamma}$ ($k$ = 0 \AA$^{-1}$). ARPES intensities are divided by the Fermi distribution function convolved with the instrumental resolution.
(b, c) ARPES (b) momentum distribution curves (MDCs) and (c) energy distribution curves (EDCs) taken from the 2D data shown in (a).
Triangle markers indicate the peak positions.
The open triangles with error bars in (c) are the energy positions of broad features. The width of the bars are explained in the text.
(d) 2D plot of the peak positions in (b, c). The bars with open triangles are the same as those in (b).
The shaded area in the left side is the projected bulk bands from ref. \cite{Baruselli16}.
Fat curves are the traces of the peak positions. These curves are copied on (a).
}
\end{figure*}

\begin{figure*}[p]
\includegraphics[width=150mm]{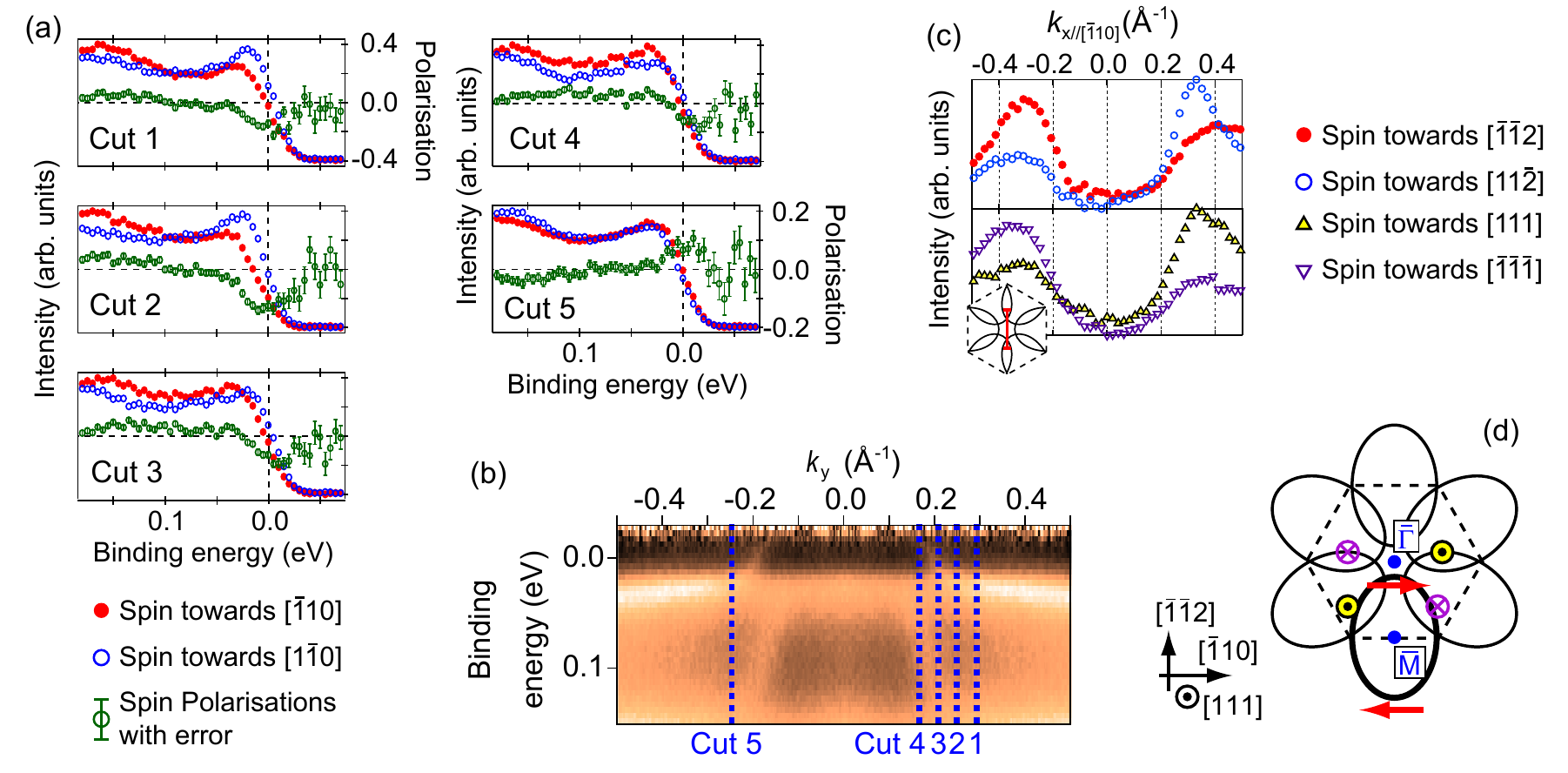}
\caption{\label{fig4} {\bf SARPES spectra} 
(a, b) SARPES spectra taken with linearly polarized photons ($h\nu$ = 26 eV) at 20 K. Detailed experimental geometries are shown in the supplementary note 1.
(a) Spin-resolved energy distribution curves (EDCs) around $k_{\rm F}$ together with spin polarisations. Errors of spin polarisation values are standard statistical errors from photoelectron counting. 
(b) Spin-resolved momentum distribution curves (MDCs) along $\bar{\Gamma}$--$\bar{\rm K}$ at $E_{\rm F}$. An inset is the schematic drawing of the Fermi contour together with the $k$ range where the spin-resolved MDCs were observed.
(c) The same as Fig. 2 (c) to indicate the positions where the spin-resolved EDCs in (a) were observed.
(d) A schematic drawing of the spin texture of the Fermi contours formed by topological surface states on SmB$_6$(111).
The arrows and circles with crosses and dots inside depict the in-plane and out-of-plane spin polarisations, respectively.
}
\end{figure*}

\newpage

\renewcommand{\thefigure}{\arabic{figure}S}
\setcounter{figure}{0}

\section*{Supplementary Information for: Non-trivial surface states of samarium hexaboride at the (111) surface}

\section*{Supplementary Note 1: Experimental geometry of angle-resolved photoelectron spectroscopy (ARPES) and spin-resolved ARPES (SARPES)}

Figure S1 shows the geometry of ARPES and SARPES setup in this work.
All of the geometries used in this work at UVSOR-III, Photon Factory, and HiSOR were nearly the same with only a small quantitative difference: photon-incident angles with respect to the normal of the electron analyser ranged from 45 to 50$^{\circ}$.
A hemispherical electron analyser with 2D electron detector detects a range of photoelectrons whose emission angles lie in the $xz$ plane in Fig. S1.
The sample was tilted to observe the 2D Fermi contour (FC), as shown in Fig. S1 (a).
The photon-incident plane for Figs. 2 and 3 in the main text is slightly ($\sim$10$^{\circ}$) away from the high-symmetry ($\bar{1}$10) plane.
Thanks to this geometry together with off-normal-incident circularly polarized photons, no symmetry operation in the photoexcitation matrix element vanishes the photoelectrons from the surface states in Figs. 2 and 3.

\begin{figure}
\includegraphics[width=80mm]{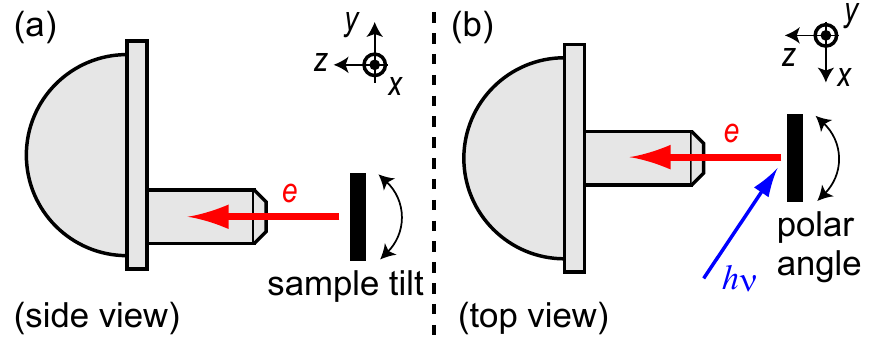}
\caption{\label{fig1s}
Schematic drawings of angle-resolved photoelectron spectroscopy (ARPES) and spin-resolved ARPES geometry.
}
\end{figure}

In the SARPES measurement, only the photoelectrons normal to the electron analyser (red arrows in Figs. S1 (a) and S1 (b)) are detected with spin separation by very low energy electron diffraction (VLEED) spin detector \cite{Okuda11}.
The tilt and azimuthal angles (sample rotation around the $z$ axis) are used to change the in-plane wave vector of each SARPES spectra shown in Fig. 4 in the main text.
{
The SARPES setup used in this work has two independent VLEED targets whose surfaces are normal to the $x$ and $y$ axes in Fig. S1, respectively \cite{Okuda15}.
For the spectra shown in Fig. 4, the latter target, resolving the spin polarization along $x$ and $z$ in Fig. S1, was used.
The $x$ orientation is parallel or anti-parallel to [1$\bar{1}$0] or [$\bar{1}\bar{1}$2] depending on the sample azimuthal angle.
The $z$ orientation is nearly parallel to the surface normal of the sample.
}
Along the polar angle, transition matrix elements are different for the positive and negative emission angles resulting in the different photoelectron intensities as well as spin polarisations \cite{Henk18}.
Such artificial asymmetry from the experimental geometry is cancelled out in the spin-resolved spectra by using the tilt angle.
For the spin-resolved spectra in Fig. S6, the polar angle (sample rotation around the $y$ axis) was used { with the other VLEED target whose surface is normal to $x$, resolving the photoelectron spins along $y$ (parallel/anti-parallel to [1$\bar{1}$0] in the geometry for Fig. S6) and $z$ (along [111]).
}

\begin{figure}[b]
\includegraphics[width=150mm]{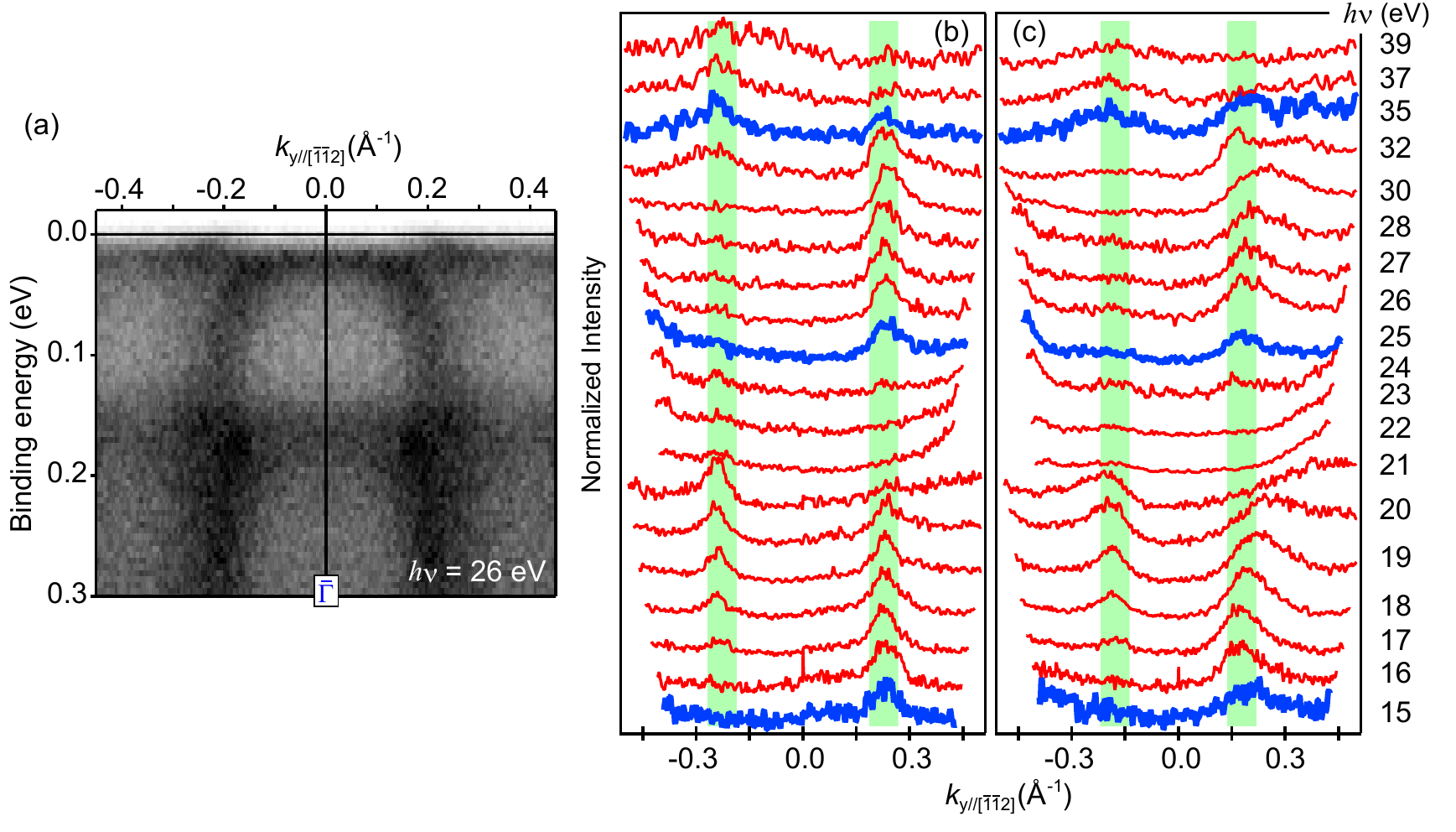}
\caption{\label{fig2s}
(a) ARPES intensity plot taken with linearly polarized photons ($h\nu$ = 26 eV) along $\bar{\Gamma}-\bar{\rm M}$ at 15 K. The ARPES intensities are symmetrized with respect to $\bar{\Gamma}$.
(b, c) APRES momentum distribution curves (MDCs) along $\bar{\Gamma}-\bar{\rm M}$ at 15 K, taken with linearly polarized photons at photon energies from 15 to 39 eV. The binding energies are at (b) the Fermi level (0 meV) and (c) 60 meV. The energy windows are 10 meV. Green fat lines are the guides to the eye.
The triangles in (c) represents the humps away from the green lines.
}
\end{figure}

\newpage

\section*{Supplementary Note 2: Photon-energy dependence of the bands around the Fermi level}

Figure S2 (a) shows the ARPES intensity plot taken with 26 eV photons, different from those shown in Figs. 2 and 3 in the main text.
It shows nearly the same dispersions of surface bands, $S1$, $S2$ and $F$, as those taken with 35 eV photons, suggesting their 2D origin.
To make the comprehensive analysis, incident-photon-energy dependent momentum distribution curves (MDCs) at the Fermi level and 60 meV (crossing $S2$) are shown in Figs. S2 (b) and S2 (c), respectively.
As guided by the fat lines, it is clearly shown that the peak positions of $S1$ show no change depending on the incident photon energies.
The case is similar for $S2$ with the peaks staying in the fat lines.
However, one can find broad features away from the fat line, indicating the $k_{\rm z}$ dispersion.
They would be from the Sm-5$d$ bulk bands.
The energy range checked here, 15 to 39 eV, corresponds to 1.1 \AA$^{-1}$ along the surface normal, assuming the inner potential of 10 eV (this is a typical value of the inner potential for the calculation of the wave vector along the surface normal).
While 1.1 \AA$^{-1}$ is $\sim$80 \% of the wavenumbers between $\Gamma$ and R in the bulk BZ, it could be reduced to 40 \% if the inversion at $\Gamma$ is assumed.
From this $k_z$ range, we cannot conclude if $S_2$ is independent from the bulk bands found above or $S_2$ itself is a part of such $k_z$ dispersion.
The additional observation with even wider $k_{\rm z}$ range is difficult, because the intensities of $S1$ and $S2$ decreases drastically in the photon energy range away from what is shown here.
It would be due to the photoexcitation cross section.
Based on this analysis, we discussed two alternative scenarios for $S_2$; as a surface resonance or a part of bulk Sm-5$d$ bands in the main text.

\begin{figure}
\includegraphics[width=150mm]{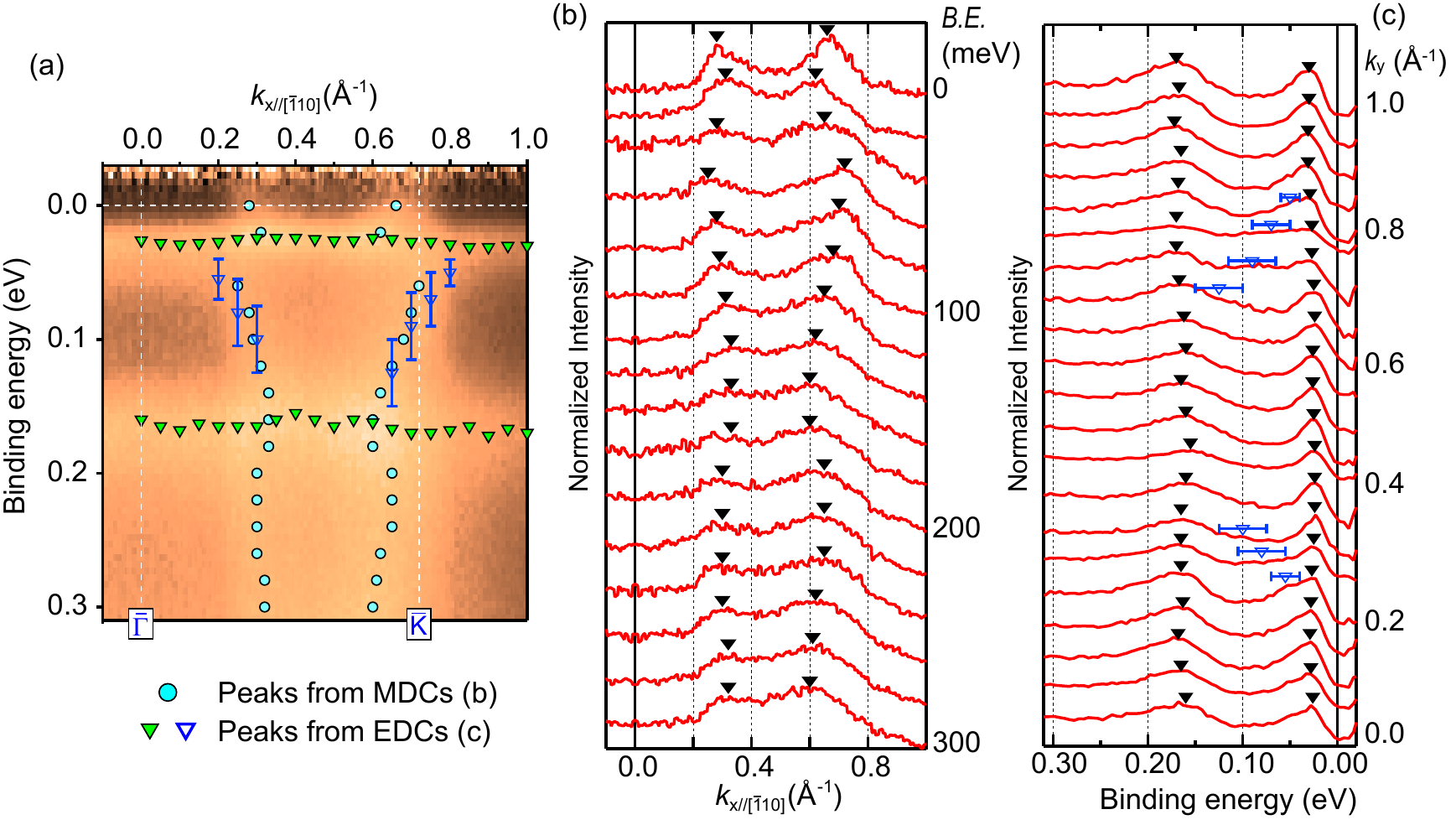}
\caption{\label{fig3s}
(a) ARPES intensity plot taken with circularly polarized photons ($h\nu$ = 35 eV) along $\bar{\Gamma}-\bar{\rm K}$ at 15 K. 
(b, c) ARPES (b) MDCs and (c) EDCs taken from Fig. S3 (a).
The filled triangles in (b, c) indicate the peak positions.
The open triangles with error bars in (c) are the energy positions of broad features. The width of the bars are explained in the text.
}
\end{figure}

\subsection{Supprementary Note 3: Band dispersion around the Fermi level along $\bar{\Gamma}-\bar{\rm K}$}

Figure S3 shows the band dispersion of SmB$_6$(111) along $\bar{\Gamma}-\bar{\rm K}$, traced by the ARPES energy and momentum distribution curves.
Similar to those along $\bar{\Gamma}-\bar{\rm M}$ (Fig. 3 in the main text), there are metallic state crossing the Fermi level at $\sim$0.25 \AA$^{-1}$ and 0.7 \AA$^{-1}$, nearly localized state around 0.03 eV, and the other dispersive bands between 0.03 to 0.18 eV, as indicated by the peaks overlaid on the 2D ARPES image (Fig. S3 (a)).

\begin{figure}
\includegraphics[width=110mm]{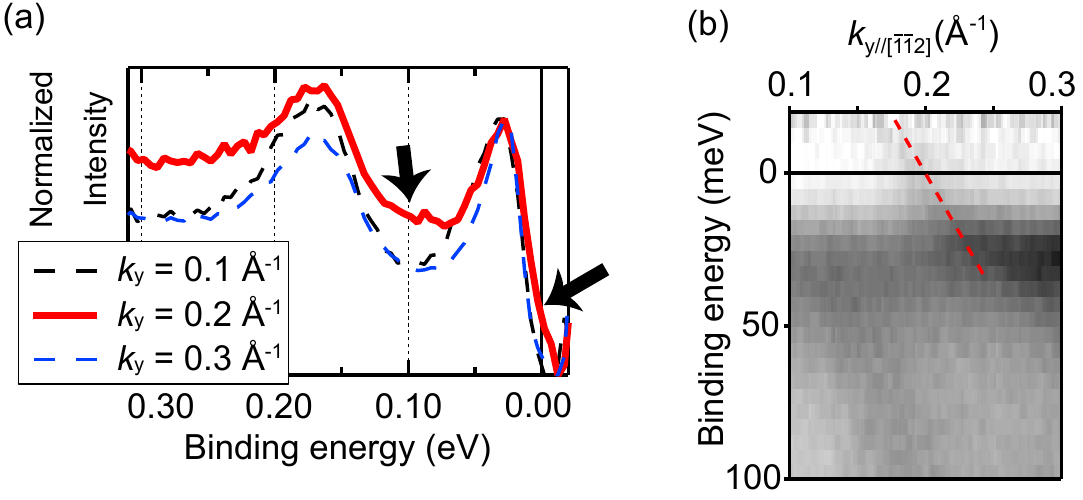}
\caption{\label{fig4s}
(a) ARPES EDCs along $\bar{\Gamma}-\bar{\rm M}$ (the same data as shown in Fig. 3 (c) in the main text) overlapped with each other.
(b) ARPES close-up image along $\bar{\Gamma}-\bar{\rm M}$ around $k_{\rm F}$, taken from the same data as Fig. 3. A dashed line is the guide of the Fermi velocity ($\sim$0.8 eV \AA).
}
\end{figure}

\subsection{Supprementary Note 4: Additional data for ARPES analysis along $\bar{\Gamma}-\bar{\rm M}$}

On the EDCs, it is difficult to trace the bands with steep dispersion, as $S1$ and $S2$ along $\bar{\Gamma}-\bar{\rm M}$.
However, they actually appear as broad features on EDCs.
To show it clearly, Fig. S4 (a) shows the EDCs at three $k_y$ points along $\bar{\Gamma}-\bar{\rm M}$; each spectrum is normalized by the peak height at $\sim$0.03 eV.
As indicated by the allows, the photoelectron intensities at the Fermi level and $\sim$0.1 eV at $k_{\rm y}$ = 0.2 \AA$^{-1}$ are higher than the others; they correspond to $S1$ and $S2$.

The Fermi velocity of $S1$ is an important parameter to be compared with those obtained by the other methods.
For a reference, we estimated the Fermi velocity.
Around the Fermi level, the dispersion of $S1$ is nearly linear as the dashed guide in Fig. S4 (b).
Based on this slope, the Fermi velocity for $S1$ along  $\bar{\Gamma}-\bar{\rm M}$ is estimated to be 0.8 eV \AA.

\subsection{Supprementary Note 5: Sizes of FCs obtained by ARPES}

\begin{figure}
\includegraphics[width=80mm]{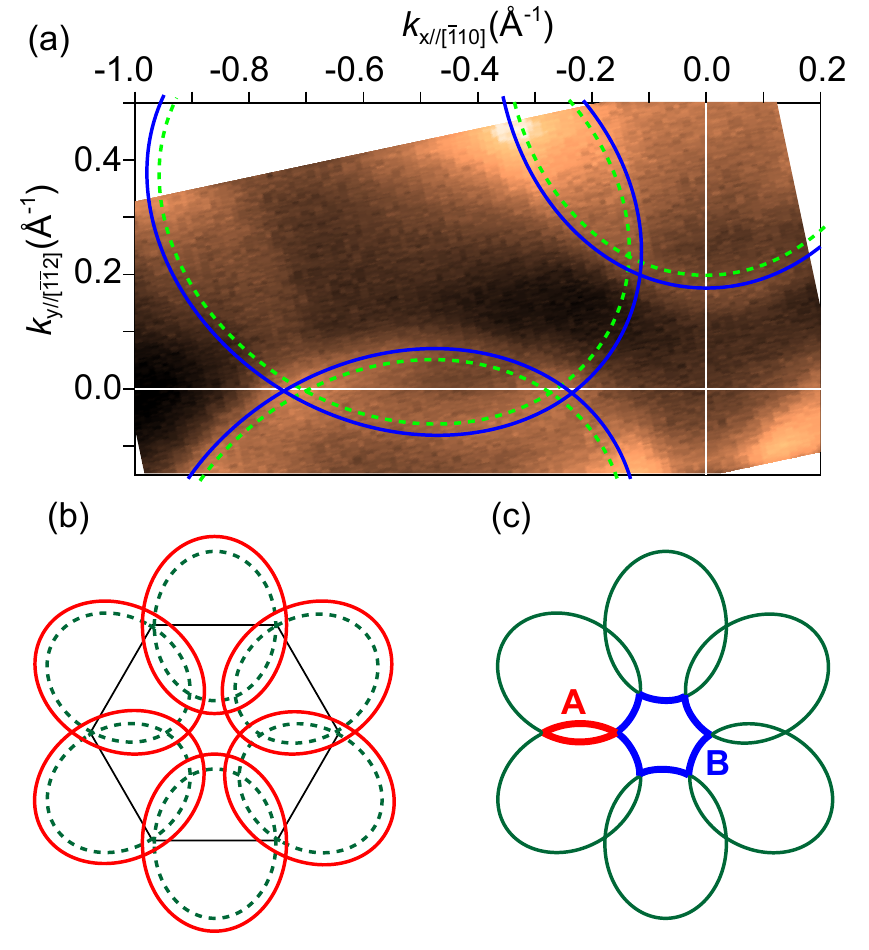}
\caption{\label{fig5s}
(a) ARPES constant-energy intensity plot at the binding energy at 70 meV (energy window of 10 meV) from the same data as Figs. 2 and 3 in the main text.
Solid curves guides the contour shape and dashed ones are FC.
(b) Virtual Fermi contours estimated by the extrapolation of S2 (thick, red curves) and the actual one from S1 (dashed curves).
(c) Another interpretation of FC. A and B correspond to the contour sizes shown in Table 1.
}
\end{figure}

The Fermi surfaces of other $R$B$_6$ materials ($R$: a rare-earth element) with the same CsCl-type lattice as SmB$_6$ are already known, such as those of CeB$_6$ \cite{Neupane15, Koitzsch16}, where the Fermi surface is the oval ones touching with each other.
The FCs of the topological surface states of SmB$_6$(111) observed here is larger than them, showing overlaps along $\bar{\Gamma}-\bar{\rm K}$. 
It means the FCs of valence-fluctuated SmB$_6$, whose valence of Sm is around 2.6, is larger than that of CeB$_6$ with trivalent Ce.
In order to examine the role of $c$--$f$ hybridisation to the size of FCs, the size of the constant-energy contour below the hybridization energy ($\sim$30 meV) is shown in Fig. S5 (a).
The size of the contour is slightly larger than that of FC ($k_y$ = 0.17 \AA$^{-1}$, $\sim$5 \% larger semi-major axis) and this change is isotropic to any orientations.
Since the dispersion of $S2$ at 100$\pm$40 meV from MDCs are nearly linear, we could extrapolate them to estimate a size of a virtual FC without $c$--$f$ hybridization, as shown in Fig. S5 (b).
The $k_{\rm F}$ shifts to 0.11 \AA$^{-1}$ (-0.08 \AA$^{-1}$ from the actual $k_{\rm F}$).
Although this shift is slightly smaller than that of SmB$_6$(001) ($\sim$0.12 \AA$^{-1}$) \cite{Denlinger16}, this shift causes even larger overlap of FCs.

The overlap of topological surface states (TSS) could cause interesting electronic phenomena.
In topological crystalline insulators (TCIs), the Lifshitz transition accompanied by a Van Hove singularity from the overlapped double topological Dirac cones is expected theoretically \cite{Liu13}.
To see such phenomena, the fine dispersion of TSS around the crossing points of them, $e.g.$ around $\bar{\rm K}$ on SmB$_6$(111), is required.
However, such fine analysis is difficult from the current data because of the limited energy resolution.
The better resolution, typically an order of 1 meV, is desirable to provide further insight into the overlapping TSS on TKI.

\begin{table}
\label{tab1}
\begin{ruledtabular}
\begin{tabular}{cccc}
 & Semi-minor axis (\AA$^{-1}$) & Semi-major axis (\AA$^{-1}$) & Area (kT) \\
\hline
Oval & 0.35 & 0.42 & 4.8 \\
A & 0.055 & { 0.21} & 0.39 \\
\hline
 & $k_{\rm F}$ // $\bar{\Gamma}--\bar{\rm M}$ (\AA$^{-1}$) & $k_{\rm F}$ // $\bar{\Gamma}--\bar{\rm K}$ (\AA$^{-1}$) & Area (kT) \\
\hline
B & 0.20 & { 0.27} & 1.6 \\
\hline
\end{tabular}
\caption{
Sizes of the FCs estimated from the TSS of SmB$_6$(111).
}
\end{ruledtabular}
\end{table}

The size of the FCs are also relevant to those observed by dHvA measurements \cite{Li14, Tan15, Hartstein17, Xiang17}.
For the comparison, the size of the oval FC is shown in Table I.
In addition, the overlap of the FCs enables the alternative interpretation of the FCs to be thin ellipsoids and a warped hexagon (A and B in Fig. S5 (c), respectively).
The sizes of them are also shown in Table I.
For the sake of comparison, the area of them are shown in the kT unit.

Although some values among them apparently agree with those observed by dHvA, $e.g.$ 0.4 kT of A with $\beta$ ($\sim$0.3 kT) and $\gamma$ ($\sim$0.4 kT) in ref. \cite{Li14}, one should be careful that it is not yet clear if such comparison could make sense.
The TSS observed here and the other ARPES works \cite{Miyazaki12, Xu13, Neupane13, Jiang13, Xu14, Hlawenka15} could, in principle, contribute to the electron transport on the crystal surface.
In contrast, the Fermi surfaces from dHvA were observed in the condition without bulk electric conduction.
Therefore, further discussion is required to understand the agreement of the FC sizes of the TSS observed by ARPES with the FS sizes by dHvA.
From experimental aspect, the dHvA measurement around the (111) surface of SmB$_6$ obtained by the similar method to this work would provide useful information for such comparison.


\subsection{Supprementary Note 6: Absence of out-of-plane spin polarization along $\bar{\Gamma}$--$\bar{\rm M}$}

Figure S6 shows the spin-resolved energy distribution curves (EDCs) observed at $k_{\rm y}$ = -0.2 \AA$^{-1}$. For this measurement, the polar angle was used (see Fig. S1 (a)).
The intense in-plane spin polarisation shown in Fig. S6 (a) (the upper panel) agrees well with those shown in Figs. 4 (a) and 4 (c) in the main text.
Such agreement independent of the experimental geometry of the SARPES measurement suggests that this spin polarisation is due to the initial states and the spin-dependent photoexcitation process (so-called final-state effect) plays no major role for the spin polarisations observed in this work.
The lower panel of Fig. S4 (a) shows that the out-of-plane spin polarisation is negligibly small.
It is natural because the $\bar{\Gamma}$--$\bar{\rm M}$ line is in the surface mirror plane.
On the mirror plane, only the spin polarisation normal to the mirror plane is allowed.

\begin{figure}
\includegraphics[width=120mm]{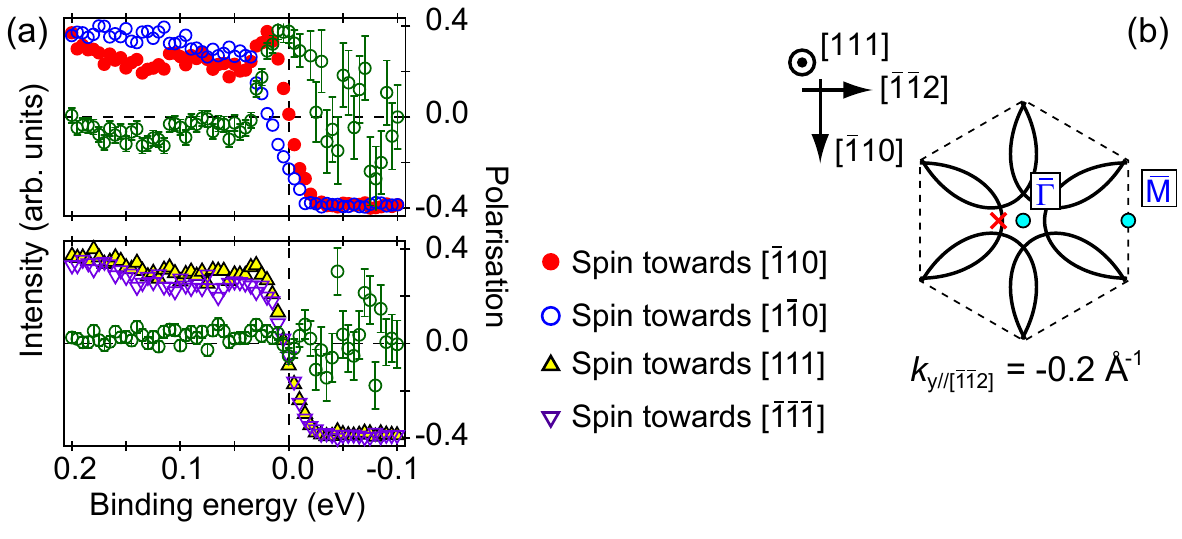}
\caption{\label{fig6s}
(a) SARPES energy distribution curves (EDCs) around $k_{\rm F}$ together with spin polarisations along the in-plane (the upper panel) and out-of-plane (the lower) orientations. Errors of spin polarisation values are standard statistical errors from photoelectron counting.
(b) A schematic drawing of the FC by the metallic surface states. A red cross indicates the position where the spin-resolved EDCs in (a, b) were measured.
}
\end{figure}

\subsection{Supprementary Discussion 1: Spin-resolved ARPES: out-of-plane spins with $C_{\rm 3v}$ surface symmetry}

\begin{figure}
\includegraphics[width=100mm]{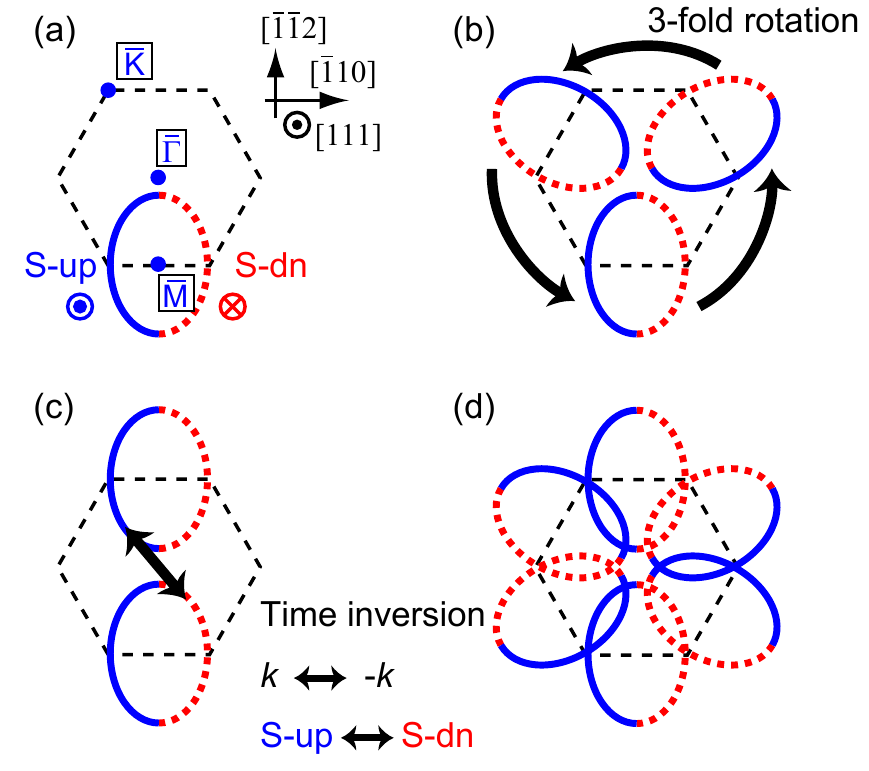}
\caption{\label{fig7s}
Schematic drawings of the out-of-plane spin-polarized FCs with symmetry operations.
(a) Single FC at a $\bar{\rm M}$ point.
(b) FCs multiplied by three-fold rotation.
(c) FCs multiplied by time inversion.
(d) FCs multiplied by the combination between (b) and (c).
}
\end{figure}

Figure S7 shows the schematic drawings of the influence of the $C_{\rm 3v}$ surface symmetry operations to the out-of-plane spin polarisations of the FCs observed in this work.
By the three-fold rotation, a FC around a $\bar{\rm M}$ point is multiplied to 3 equivalent FCs without any modification to the out-of-plane spin polarisations, as shown in Fig. S7 (b).
On the other hand, the time-inversion operation multiplies a FC by changing the sign of the wave vector $k$ and spin polarisation, as shown in Fig. S7 (c).
The combination of the three-fold rotation and time inversion results in the 6 FCs as shown in Fig. S7 (d).
It should be noted that the out-of-plane spin polarisation along $\bar{\Gamma}$--$\bar{\rm K}$ is not cancelled out.
This spin texture is neither violated by the surface mirror plane, ($\bar{1}$10); all the out-of-plane spins on the FCs changes their signs with respect to the mirror operation from the left to the right sides.
The cases are the same for the other two surface mirror planes.

\subsection{Supprementary Discussion 2: Detailed analysis of spin-resolved MDCs}

In order to discuss the topological order from the surface states at $E_{\rm F}$, it is enough to exclude the doubly spin-degenerate state, as discussed in the main text.
Here, as an additional information, we perform a semi-quantitative analysis of the observed spin-resolved MDCs.

Figures S8 (a, b) are for the MDCs at the Fermi level along $\bar{\Gamma}$--$\bar{\rm K}$.
As shown in Fig. S8 (a), the $I_{tot}$ spectrum along $\bar{\Gamma}$--$\bar{\rm K}$ has asymmetric peak shape, which makes the analysis along this direction difficult.
However, this feature is not only observed by SARPES but also by conventional ARPES, as the red circular markers in Fig. S8 (a).
The slight shift of the peak positions of $I_{tot}$ from SARPES (the solid line in Fig. S8 (a)) would be due to the wider energy resolution and/or the acceptance angle.
To understand it more in detail, we fitted the conventional ARPES MDC by two Gaussians and a third broad feature as indicated by the thin dashed line in Fig. S8 (a).
The peak positions of the Gaussians, $k_{//[\bar{1}10]}$ = 0.3 and 0.7 \AA$^{-1}$, are consistent with the metallic surface states shown in Fig. S3.
The origin of the broad background would be due to the FCs lying close to the $\bar{\Gamma}$--$\bar{\rm K}$ line.
Actually, the MDC at the Fermi level normal to $\bar{\Gamma}$--$\bar{\rm K}$, as shown in Fig. S8 (b), indicates two separate peaks corresponding to such FCs, with the finite intensity at $k_{//[11\bar{2}]}$ = 0 \AA$^{-1}$ from their tail.
Therefore, the asymmetric shape of the spin-resolved MDC shown in Fig. 4 (b) in the main text would be also derived from the overlap of such background and the peak at 0.3 \AA$^{-1}$.

\begin{figure}
\includegraphics[width=150mm]{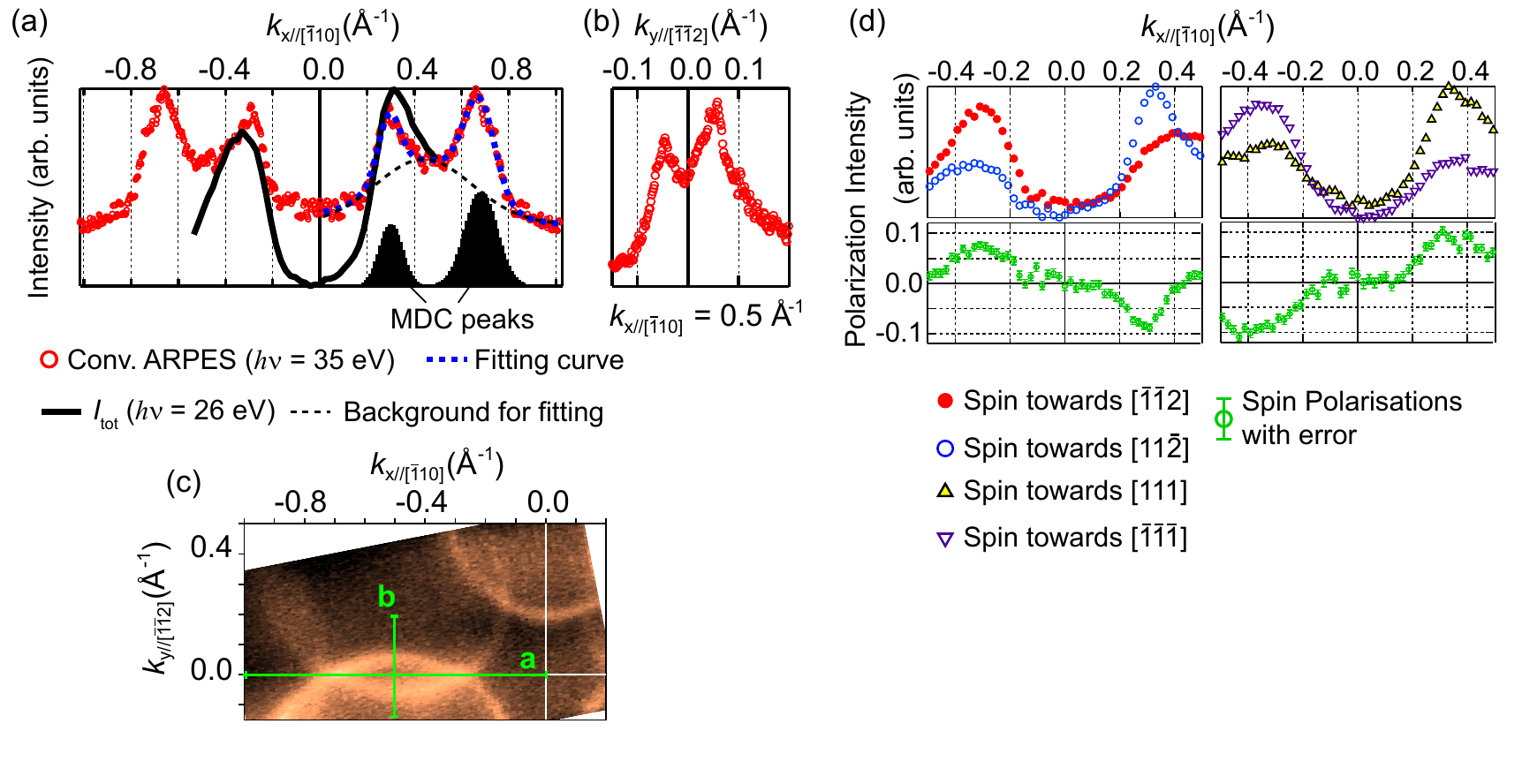}
\caption{\label{figSRMDC}
(a) Spin-integrated MDCs along $\bar{\Gamma}$--$\bar{\rm K}$ taken at $h\nu$= 35 and 26 eV at the Fermi level (the same data as Fig. S3 (b) and Fig. 4 (b), respectively). Dashed curves are the fitting curve and its background (see text for details).
The MDC at $h\nu$= 35 eV is mirrored with respect to $\bar{\Gamma}$.
(b) Spin-integrated MDC taken along [11$\bar{2}$] at $k_{//[\bar{1}10]}$ = 0.5 \AA$^{-1}$.
(c) ARPES Fermi contour to indicate the region where the MDCs are measured for (a) and (b).
(d) SARPES MDCs and spin polarizations measured along $\bar{\Gamma}$--$\bar{\rm K}$ at the Fermi level ($h\nu$= 26 eV).
Errors of spin polarisation values are standard statistical errors from photoelectron counting. 
}
\end{figure}

Fig. S8 (d) is the SARPES MDCs and the corresponding spin polarizations of the spin-resolved MDCs. 
The peaks of the in-plane spin polarizations are clearly around 0.3 \AA$^{-1}$, suggesting that this polarization corresponds to the surface states there.
The out-of-plane ones does not have such clear peak position.
Instead, it has broad intensities at $|k_{//[\bar{1}10]}| >$ 0.2 \AA$^{-1}$.
It would be because both the FC at 0.3 \AA$^{-1}$ and the background from the neighbouring FCs are spin polarized along the out-of-plane orientation.
At first glance, the spin polarization from the neighbouring FCs might appear to be cancelled out.
However, since the $\bar{\Gamma}$--$\bar{\rm K}$ line is not the mirror plane, the intensities from the neighbouring FCs could be different, as shown in Fig. S8 (b).
Therefore, finite spin polarization from them is expected.
Because of such complicated components, the quantitative analysis of the spin-resolved MDCs along $\bar{\Gamma}$--$\bar{\rm K}$ are quite difficult.
However, from the spin polarizations in Fig. S8 (d), one can see the FC from the surface states has both in-plane and out-of-plane spin components, as stated in the main text.
For the quantitative analysis, one needs wider wavevector range as well as the different incident photon energies, polarizations, and the spin polarization to the other orientations.
However, such quantitative determination of the photoelectron spin polarization is not the focal point of this research.
Note that the spin polarization modulation due to the interference of the photoelectron wavefunctions \cite{Dil15, Yaji17} is also out of the focus of this research.

\subsection{Supprementary Discussion 3: Comparison with the ARPES results on SmB$_6$(001)}

\begin{figure}
\includegraphics[width=100mm]{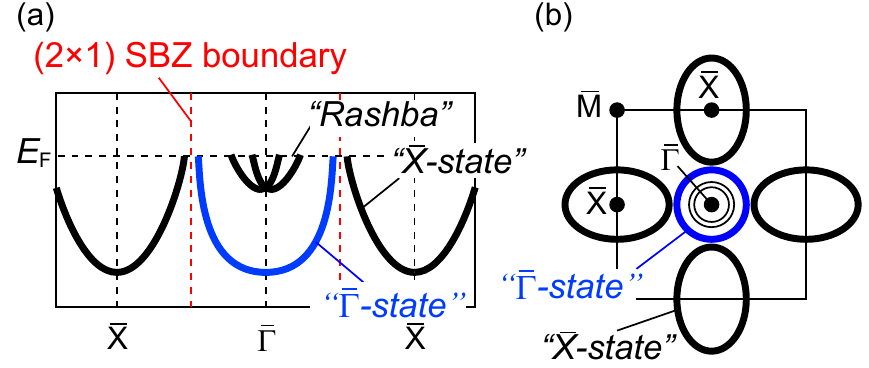}
\caption{\label{fig8s}
Schematic drawings here are based on what were observed in high-resolution ARPES studies \cite{Miyazaki12, Xu13, Neupane13, Jiang13, Xu14, Hlawenka15}.
Assignments are according to ref. \cite{Hlawenka15}.
The thin double-circle around $\bar{\Gamma}$ in (b) is the ``Rashba'' band in (a).
}
\end{figure}

We've provided the smoking-gun evidence on the non-trivial topological order of SmB$_6$ based on our spin-integrated and spin-resolved ARPES measurements.
Therefore, the surface states on the (001) surface, which has been under debate \cite{Miyazaki12, Xu13, Neupane13, Jiang13, Xu14, Hlawenka15} for several years, should also indicate the non-trivial topological order in principle, because the topological order of materials does not depend on the surface orientation but on its bulk electronic structure and its parities.
This problem, the controversial topological assignments on SmB$_6$(001), can be solved by considering the interpretation of the ``$\bar{\Gamma}$-state'' observed in ref. \cite{Hlawenka15}.
Figure S9 shows the schematic drawings of the surface band structure observed in earlier ARPES studies \cite{Miyazaki12, Xu13, Neupane13, Jiang13, Xu14, Hlawenka15}.
In ref. \cite{Hlawenka15}, the ``$\bar{\Gamma}$-state'' band is assigned as a folding of the surface bands surrounding $\bar{X}$ (``$\bar{X}$--$state$'') with respect to the (2$\times$1) surface Brillouin zone boundaries.
However, the slope of the ``$\bar{\Gamma}$-state'' band is different from that of its counterpart, as shown in Fig. 2 (c) of ref. \cite{Hlawenka15}, suggesting that this state would be a different, individual metallic state.
If this state is not an umklapp state but the new state independent of the ``$\bar{X}$--$states$'', there are three closed FCs on the SmB$_6$(001) surface; one is around $\bar{\Gamma}$ (by the ``$\bar{\Gamma}$-state'' state) and two are around two $\bar{\rm X}$ points (``$\bar{X}$--$states$''), as depicted in Fig. S9 (b).
Based on this interpretation, one can conclude that there are three closed spin-polarised FCs on both the (111) and (001) surfaces of SmB$_6$, which indisputably indicates its non-trivial topological order.
The ``Rashba'' state observed in ref. \cite{Hlawenka15} (see Fig. S9 (a)) plays no role in the topological order of the material because it always forms even numbers (2, in this case) of closed FCs around the surface TRIMs.
It should be noted that this interpretation is just a possibility based on known results.
The detailed origin of the ``$\bar{\Gamma}$-state'' band would be elucidated by future ARPES and SARPES works with varying incident photon energies and polarisations.

{
\subsection{Supprementary Discussion 4: Possible origins of electronic states observed around $E_{\rm F}$ in this work}
Historically, the two mechanisms to form the surface electronic states have been discussed \cite{Desjonqueres96}.
In a localized-bond picture, surface states can be induced by the surface atomic structure independently from the bulk bands, as the dangling-bond states.
The other case is a perturbed-bulkband picture. 
Surface states are derived from bulk Bloch states, which are perturbed by the truncation of the periodic potential at the surface and consequently localized in near-surface region.
In this case, the surface band tends to show the nearly parallel dispersion and similar orbital character to the ``mother'' bulk bands.
Quantum-well states confined in the surface layers, as two-dimensional electron gas states on SrTiO$_3$ \cite{Son09}, also belong to this group.
In this context, the surface band $S1$ in this work clearly belongs to the latter case, closely related to the bulk Sm 5d and 4f bands, as evidenced by its dispersion similar to what is expected to the bulk Sm 5d and 4f bands around $E_{\rm F}$.
If $S2$ and $F$ are also the surface-derived ones, they would also belong to the latter case.

In order to explain the surface electronic structure of SmB$_6$, some theoretical models were discussed.
Zhu \textit{et al.} proposed a metallic surface state formed by the combination between the surface polarity and boron dangling bonds at the surface \cite{Zhu13}.
Based on the classification above, it is not likely a case for $S1$, since the dispersion of the dangling-bond state is completely different from the bulk bands in most cases.
Note that a possible surface-band formation based on this mechanism is not excluded.
Such state can also be formed away from $E_{\rm F}$ and it would be the case for SmB$_6$.
Actually, the energies of such surface state was calculated to be sensitive to the surface termination condition \cite{Zhu13}.
The other mechanism was proposed in ref. \cite{Hlawenka15} as ``many-body resonance''.
It was claimed to be originate from the energetic shift of the bulk bands in the surface atomic layers.
The ``many-body resonance'' has similar orbital character and dispersion to its bulk counterparts, but independent from them due to the different boundary condition in the surface atomic layers.
Therefore, it is a member of the group of ``perturbed bulk bands'' and would be a likely case to describe the characteristics of the current electronic states.

It should be noted that all of the proposed mechanisms are compatible to the topological classification.
For example, a quantum-well-like state can be a TSS at the same time as simulated in HgTe thin films \cite{Luo11}.
The classification procedure merely counts the number of Fermi contours formed by the metallic surface states.
As far as the candidate surface state is localized in the surface layers and out of the projected bulk bands, topological classification is always valid, based on the present knowledge.
We found no reason to exclude the shifted bulk bands localized in surface atomic layers.
Nonetheless, the limitation of the topological classification is discussed in the main text.
}

\end{document}